\documentclass[superscriptaddress,aps,pra,twocolumn,showpacs,nofootinbib,longbibliography,notitlepage]{revtex4-1}
\usepackage{etex}
\usepackage{amsmath,amssymb,amsthm}
\usepackage[colorlinks=true,citecolor=blue,urlcolor=blue]{hyperref}
\usepackage[pdftex]{graphicx}
\usepackage{times,txfonts}
\usepackage{braket}
\usepackage{color}
\usepackage{natbib}
\usepackage{amsmath,blkarray}
\usepackage{mathtools}
\usepackage{latexsym}
\usepackage{tabularx, booktabs}
\usepackage{graphics,epstopdf}
\usepackage{graphicx}
\usepackage{float}
\usepackage{graphicx}
\usepackage{amsfonts}
\usepackage{subcaption}
\usepackage{color,soul}

\newcommand{\be}{\begin{equation}}
\newcommand{\ee}{\end{equation}}
\newcommand{\ba}{\begin{eqnarray}}
\newcommand{\ea}{\end{eqnarray}}

\begin{document}
\title{Activation of hidden nonlocality using local filtering operations based on CGLMP inequality}
\author{ Asmita Kumari }
\email{asmita.physics@gmail.com}
\affiliation{S. N. Bose National Centre for Basic Sciences, Block JD, Sector III, Salt Lake, Kolkata 700106, India}

\begin{abstract}
Entanglement is necessary but not sufficient to demonstrate nonlocality as there exist local entangled states which do not violate any Bell inequality. In recent years, the activation of nonlocality (known as hidden nonlocality) by using local filtering operations has gained considerable interest. In the original proposal of Popescu [\href{https://link.aps.org/doi/10.1103/PhysRevLett.74.2619}{Phys. Rev. Lett.  74, 2619 (1995)}] the hidden nonlocality was demonstrated for the Werner class of states in $d \geq 5$. In this paper, we demonstrate the hidden nonlocality for a class of mixed entangled states (convex mixture of a pure state and color noise) in an arbitrary $d$-dimensional system using suitable local filtering operations. For our demonstration, we consider the quantum violation of Collins-Linden-Gisin-Masser-Popescu (CGLMP) inequality which has hitherto not been considered for this purpose.  We show that when the pure state in the aforementioned mixed entangled state is a maximally entangled state, the range of the mixing parameter for revealing hidden nonlocality increases with increasing the dimension of the system. Importantly, we find that for $d \geq 8$, hidden non-locality can be revealed for the whole range of mixing parameter. Further, by considering another pure state, the maximally CGLMP-violating state, we demonstrate the activation of nonlocality by using the same local filtering operation.

 %Entanglement provides the necessary but not sufficient conditions to generate nonlocal correlations through the violation of Bell inequalities. There exist entangled states that do not violate any Bell inequality. Because of the importance of nonlocal correlations in quantum information theory, the activation of nonlocality known as hidden nonlocality in entangled state admitting local models has been extensively studied. In this paper, we study the phenomenon of revealing nonlocality using local filtering operation for arbitrarily high dimensional $(d>2)$ quantum systems that can be checked through the violation of the local bound of Collins-Linden-Gisin-Masser-Popescu (CGLMP) inequality. By considering single copy of $d$-dimensional two-qubit mixed entangled states we show that range of mixing parameters showing hidden non-locality increases with an increase in dimension. For $d \geq 8 $, we found that hidden non-locality is observed for whole range of the mixing parameter of the mixed entangled state. We have provided numerical results for $d=100$.

\end{abstract}

\maketitle

\section{Introduction}
Bell's theorem \cite{Bell1964} is one of the most remarkable discoveries of quantum theory. This no-go theorem elegantly discriminates the quantum theory from the local classical theories by demonstrating that all the predictions of quantum theory cannot be reproduced by local realist models. Such a feature widely known as quantum nonlocality and is commonly demonstrated through the quantum violation of suitable Bell inequalities. Note that, the entanglement \cite{Schrödinger1935,epr1935,horodecki09} between specially separated quantum systems is necessary for demonstrating nonlocality. However, entanglement is not sufficient to generate a nonlocal quantum correlation. There exist entangled states which admit local realist models and hence do not violate any Bell inequality.  It remains a challenging problem in higher dimensional systems to find the connection between entanglement and nonlocality.

% Entanglement \cite{Schrödinger1935,epr1935,donald02,horodecki09}, defined as the inseparability of the physical systems is a quantum resource that plays a crucial role in the development of quantum information processing \cite{Ekert1991,Bennett92,Bennett93}.

In the last few decades, much effort has been put forward to activate nonlocality (commonly known as hidden nonlocality) for various kinds of entangled states admitting local models. In his pioneering paper, Popescu \cite{Popescu1995} first demonstrated the quantum violation of Bell-Clauser-Horne-Shimony-Holt (CHSH) inequality by applying local filters on the Werner states \cite{Werner1989} admitting local model. Such an activation of nonlocality by applying filtering operation on local entangled state is valid in $d \geq 5 $. By using different class of state Gisin \cite{GISIN1996}, demonstrated a nonlocality activation protocol for a suitably chosen two-qubit local entangled state through the quantum violation of CHSH inequality \cite{HORODECKI1995340,clause1969}. Later, local filtering with projective measurement was generalized for positive operator value measures in \cite{Brunner2013}. An alternative route of activating nonlocality is also introduced in \cite{peres1996,aditi2005,pala2012,bancal2011} by using multiple copies of the entangled state while each of the copies admits a local model.  

The present papers also concerns the activation of nonlocality (or revealing hidden nonlocality) of $d$-dimensional quantum systems using local filtering operations. Most works towards this direction are limited to $2$-dimensional local quantum systems and demonstrated through the violation of CHSH inequality \cite{Popescu1995,GISIN1996, Brunner2013,frank2002, Pramanik2019,ming2017} as it provides the necessary and sufficient conditions. The original work of  Popescu \cite{Popescu1995} was for high-dimensional Werner state in $d \geq 5 $. But, activation of nonlocality for other class of high-dimension local entangled states is less explored. Hirsch et. al. \cite{Brunner2013} strengthen the argument in \cite{Popescu1995} for POVMs by considering qutrit-qubit and qutrit-qutrit entangled state admitting local models. However, the generalized argument to demonstrate hidden nonlocality using suitable Bell inequalities for any arbitrary dimensional system remains unexplored. We note here that, for higher outcome systems, the well-known Collins-Gisin-Linden-Massar-Popescu (CGLMP) \cite{Collins2002,pin2016} inequalities provide necessary and sufficient conditions to demonstrate nonlocality. Hence, it could be an interesting line of study to explore the hidden nonlocality for an arbitrary high-dimensional system using suitable local filtering based on the quantum violation of CGLMP inequality.

In this work, we demonstrate the hidden nonlocality for arbitrary high-dimensional bipartite local entangled states using local filtering operation through the quantum violation of CGLMP inequality.  We consider a class of states for our study is given by \cite{cabello05}
\begin{eqnarray}
\label{pd}
\rho_{d} = q |\psi_d \rangle \langle \psi_d | + (1-q) | 0 \rangle \langle 0 | \otimes \frac{\mathbb{I}_d}{d}
\end{eqnarray}
where $q$ ($0 < q \leq 1$) is the mixing parameter and $|\psi_d\rangle\in \mathcal{C}^{d}\otimes \mathcal{C}^{d}$ is a pure state. By first taking the mixed entangled state where $|\psi_d\rangle$ is a  maximally entangled state, we show that the range of mixing parameter showing hidden non-locality increases with increasing the dimension of the system. Importantly, we demonstrate that for $d \geq 8$, the hidden non-locality can be revealed for any non-zero value of the mixing parameter. We note that, except for $d=2$, CGLMP inequality is maximally violated by a non-maximally entangled state (known as maximally CGLMP-violating state) \cite{Acin2002}. Further, we explore the hidden nonlocality when $|\psi_d\rangle$ in Eq. (\ref{pd}) is a maximally CGLMP violating state.  We observe that the hidden nonlocality can be demonstrated for a comparatively wider range of mixing parameter compared to the former case while $|\psi_d\rangle$ is a maximally entangled state.

This paper is organized as follows. In Sec. II, we focus on preliminaries which includes a brief discussion on the quantum violations of CHSH inequalities and CGLMP inequalities. We discuss the effect of local filtering on $2$-dimensional mixed entangled state in Sec. III.  In Sec. IV, we generalize the application of local filtering operation for $d$-dimensional bipartite quantum system using CGLMP inequality. Finally, in Sec V, we discuss our results.

\section{Preliminaries}
Before presenting our results on activating nonlocality by using local filtering operations, let us first discuss the range of mixing parameter for the state in Eq. (\ref{pd}) for which the quantum violation of CHSH and CGLMP inequalities is obtained.
\subsection{Quantum violation of CHSH inequality}
The CHSH inequality is the simplest Bell's inequality defined in two-party, two-measurement, and two-outcome per measurement scenario \cite{clause1969}. In the CHSH scenario, two space-like separated parties, Alice and Bob perform measurements of two dichotomic observable  $(A_{1}, A_{2})$ and $(B_{1}, B_{2})$ respectively. The CHSH inequality valid for any local theory can be written as
\begin{eqnarray}
\label{slocal}
 \langle A_1  B_1 \rangle + \langle A_1  B_2 \rangle + \langle A_2  B_1 \rangle-\langle A_2  B_2 \rangle \leq 2
\end{eqnarray}
By considering that Alice and Bob share a two-qubit state (putting $d=2$ in Eq. (\ref{pd})) is of the form
\begin{eqnarray}
\label{pr}
\rho_{2} = q |\psi_2 \rangle \langle \psi_2 | + (1-q)|0 \rangle \langle 0 | \otimes  \frac{\mathbb{I}_2}{2}
\end{eqnarray}
where $|\psi_2 \rangle =  \frac{1}{\sqrt{2}}(|00 \rangle +  |11 \rangle$ is maximally entangled two-qubit state. 

For a suitable choice of observable, the maximum quantum value of CHSH inequality for the state in Eq. (\ref{pr}) is obtained to be $2 \sqrt{2}q$. It puts restriction on $q$ ($q > \frac{1}{\sqrt{2}} = 0.707$) for the violation of CHSH inequality. In the range of $0 < q \leq 0.707 $ even if the state $\rho_2$ is entangled, it does not violate CHSH inequality. Following Popencu's \cite{Popescu1995} idea, Gisin\cite{GISIN1996} demonstrated that using local filtering the above range can be made narrower, i.e., for a lower value of mixing parameter CHSH inequality can be violated thereby revealing the hidden nonlocality.

\subsection{Quantum violation of CGLMP inequality}
In CGLMP scenario, Alice performs the measurement of two observable $A_1$ and $A_2$ and Bob performs $B_1$ and $B_2$. Each of the measurements produces  $d$ outcome ($0,1,2,....,d-1$). The CGLMP inequality was derived as  \cite{Collins2002,pin2016}
\begin{widetext}
\begin{eqnarray}
\label{cgl}
(I_{d})_{L} &=&  \bigg|\sum^{\left\lfloor \frac{d}{2}\right\rfloor -1}_{k = 0} \bigg( 1 - \frac{2 k}{d-1}     \bigg) \bigg[ P(A_1 = B_1 +k) + P(B_1 = A_2+k+1) + P(A_2 = B_2+k)+ P(B_2 = A_1+k) \\ \nonumber && -\bigg( P(A_1 = B_1 -k-1) + P(B_1 = A_2-k) + P(A_2 = B_2-k-1)+ P(B_2 = A_1 -k-1) \bigg)\bigg] \bigg|  \leq 2
\end{eqnarray}
\end{widetext}
which is valid for any local theory. Here, the subscript $L$ denotes the local and $ P(A_a = B_b +k) $ denotes the probabilities of the outcomes of Alice's measurement $A_a$ and Bob's measurement $B_b$ ($a,b=1,2$) that differ by $k$ mod $d$ as
\begin{eqnarray}
\label{prob}
P(A_a = B_b +k) = \sum^{d-1}_{j=0} P(A_a =j, B_b = j+k \  \ mod \  \ d)
\end{eqnarray}

If the shared state between Alice and Bob is 
\begin{eqnarray}
\label{maxent}
|\psi_{d} \rangle = \frac{1}{\sqrt{d}}\sum^{d-1}_{j=0}|j \rangle_{A}\otimes |j \rangle_{B} 
\end{eqnarray}
and the non-degenerate eigenvectors of measuring operators $A_a$ ($a = 1, 2$) of Alice and $B_b$ ($b = 1, 2$) of Bob of the form
\begin{eqnarray}
\label{ma}
|k \rangle_{A_a} = \frac{1}{\sqrt{d}}\sum^{d-1}_{j=0}\exp(i \frac{2 \pi}{d}j(k + \alpha_a))|j \rangle_{A} 
\end{eqnarray}
and
\begin{eqnarray}
\label{mb}
|l \rangle_{B_b} = \frac{1}{\sqrt{d}}\sum^{d-1}_{j=0}\exp(i \frac{2 \pi}{d}j(-l + \alpha_a))|j \rangle_{B} 
\end{eqnarray}
respectively, with
\begin{eqnarray}
\label{v1}
 \alpha_1= 0, \ \ \ \alpha_{2} = 1/2,\ \ \  \beta_{1} = 1/4 \ \ \  and  \ \ \   \beta_{2} = -1/4.
\end{eqnarray}
Then, the joint probability can be obtained by using quantum Fourier transformation \cite{Nielsen2010QuantumCA} is given by
\begin{eqnarray}
\label{pp}
  P(A_a = k, B_b = l) = \frac{1}{d^3 }  \bigg[ \frac{1}{2 \sin^2{[\pi(k-l + \alpha_a +\beta_b)/d]}} \bigg] 
\end{eqnarray}

By using Eq. (\ref{pp}) and by putting the values of $\alpha_a$ and $\beta_b$ the optimal quantum value of CGLMP functional   is obtained as
\begin{eqnarray}
\label{cgqq}
(I_{d})_{Q}^{opt} &=& 4 d \sum^{\left\lfloor \frac{d}{2}\right\rfloor -1}_{k = 0} \bigg( 1 - \frac{2 k}{d-1}     \bigg) (p_k - p_{-(k+1)} )
\end{eqnarray}
where, 
\begin{eqnarray}
p_c =  P_{Q}(A_1 = B_1+ c ) = \frac{1}{2 \sin^2{[\pi(c + 1/4)/d]}}
\end{eqnarray}
where $c \in (k, k+1)$ is an integer which denotes that the probability of $A_a$ and $B_b$ differ by constant integer $c$. %\textcolor{red}{Note that, appropriate measurement settings and maximally entangled state in $d$-dimension provide the necessary condition for the violation of CGLMP inequality. }

Let us now consider that Alice and Bob share a mixed entangled state $\rho_d$ in Eq.(\ref{pd}) where $|\psi_d \rangle$ is a maximally entangled state. In such a case, the quantum value of the CGLMP functional gives
\begin{eqnarray}
\label{cgll}
I'_{d,Q} = q I_{d,Q}
\end{eqnarray}
which in turn imposes a lower bound on the mixing parameter $q$ for the violation of CGLMP inequality is given by
\begin{eqnarray}
\label{cgll1}
q > \frac{ 2}{(I_{d,Q})^{opt}}
\end{eqnarray}
where $(I_{d,Q})^{opt}$ is the optimal value of $I_{d,Q}$. It is demonstrated that for $d \rightarrow \infty $,  nonlocality can be observed when $q> 0.673$ \cite{Collins2002}. Hence, within the range $0 < q \leq 0.673 $ the state $\rho_d$ admits the local model as there is no violation of CGLMP inequality. 

Instead, if one takes the pure state $|\psi_d \rangle$ in $\rho_d$ to be maximally CGLMP violating state the range of mixing parameter is derived \cite{fonseca18,roy20} as $ 0.637 < q \leq 1$ for $d=10$.

We demonstrate that for both the cases where $|\psi_d \rangle$ in Eq.(\ref{pd}) is  maximally entangled state and maximally CGLMP violating state, the local filtering can reveal the hidden nonlocality within that range of mixing parameters admitting the local model.  However, the ranges are different in those two cases, as derived in our work.

\section{Hidden non-locality in $2$-dimensional mixed entangled state} 

Before proceeding to demonstrate our results, let us briefly discuss the activating the nonlocality using local filtering operation for a two-qubit mixed entangled state in Eq. (\ref{pr}). For this, we take a specific form \cite{Brunner2013} of local filtering operators for Alice and Bob are the following.  
\begin{eqnarray}
F_A = \xi | 0 \rangle \langle 0 | +  | 1 \rangle \langle 1 |;  \ \  \  \  \  \ F_B = \delta | 0 \rangle \langle 0 | +  | 1 \rangle \langle 1 | 
\end{eqnarray}
 Alice and Bob apply these local filtering operations on the respective subsystems of the shared state. Here, $0\leq \xi, \delta \leq 1$ with  $\delta =\frac{\xi }{\sqrt{q}}$. After the operation of local filters $F_A$ and $F_B$, the shared state $\rho_2$ in Eq.(\ref{pr}) transforms as
\begin{eqnarray}
\rho^F_{2} &=&   \frac{1}{N_2} (F_A \otimes F_B) \rho_{2}  (F_A \otimes F_B)^{\dagger} \\ \nonumber &=&  \frac{1}{N_2} \bigg[q  |\psi_2 \rangle \langle \psi_2 | + \sqrt{\frac{q}{2}}( \xi^2-  \sqrt{q}) \bigg( |\psi_2 \rangle \langle 0 0 | +| 0 0 \rangle \langle \psi_2| \bigg)  \\ \nonumber &&   + \frac{(1-q) \xi^2}{2}  | 0 1 \rangle \langle 0 1 |    +  \frac{\xi ^4-2 \xi ^2 q^{3/2}+q^2}{2 q}| 0 0 \rangle \langle 0 0 | \bigg]
\end{eqnarray}
where, $N_2 = \frac{1}{2} \left(q+(1-q)\xi ^2+\frac{\xi ^4}{q}\right)$ is the normalization constant. This filtered state  $\rho^F_{2}$ violates CHSH inequality for a wider range than the unfiltered state. The CHSH violation is obtained for the range of $ 0.665 < q \leq 1$ at $\xi = 0.79 $ compared to the range $ 0.707 < q \leq 1$ obtained for the unfiltered state. Hence, the action of the local filtering operator turns the local state into a nonlocal state within the range $0.665 < q <0.707$. Similar results of revealing hidden nonlocality using local filtering operations in $2$-dimensional quantum systems are discussed in \cite{GISIN1996, Brunner2013,frank2002, Pramanik2019,ming2017}. We now proceed to reveal hidden nonlocality for the $d$-dimensional system through the quantum violation of CGLMP inequality which has not hitherto been explored.

\section{Hidden non-locality in $d$-dimensional mixed entangled state}
We take the mixed entangled state as in Eq. (\ref{pd}) and the local filtering operators of Alice and Bob are of the form
\begin{eqnarray}
\label{df}
F_A = \xi | 0 \rangle \langle 0 | + \sum^{d-1}_{j=1} | j \rangle \langle j |; \ \ \ F_B = \delta | 0 \rangle \langle 0 | +  \sum^{d-1}_{j=1} | j \rangle \langle j | \\ \nonumber
\end{eqnarray}
 respectively \cite{Mat2020}. Alice and Bob apply the above local filtering operations on their respective subsystems. As mentioned, we consider two different forms of $|\psi_{d}\rangle\in\mathcal {C}^{d}\otimes\mathcal {C}^{d} $ in Eq. (\ref{pd}), viz., the maximally entangled state and the maximally CGLMP-violating state.
\subsection{ When $|\psi_d\rangle$  is maximally entangled state}

For the mixed state $\rho_{d}$ with $|\psi_d\rangle$ is the maximally entangled state as in Eq. (\ref{maxent}), the application of local filtering operation (in Eq.(\ref{df})) transforms $\rho_{d}$  as
\begin{widetext}
\begin{eqnarray}
\label{dfr}
\rho^F_{d} &=&  \frac{1}{  N_d} (F_A \otimes F_B) \rho_{d}  (F_A \otimes F_B)^{\dagger} \\ \nonumber &=&   \frac{1}{d  N_d} \bigg[q \bigg( \sum^{d-1}_{j=0}  |j \rangle   |j \rangle \bigg) \bigg( \sum^{d-1}_{j=0}   \langle j |  \langle  j | \bigg)    + \sqrt{q}( \xi^2-  \sqrt{q}) \bigg[ \bigg( \sum^{d-1}_{j=0}  |j \rangle   |j \rangle \bigg) \langle 0 0 |   +| 0 0 \rangle \bigg( \sum^{d-1}_{j=0}   \langle j |  \langle  j |  \bigg) \bigg]  \\ \nonumber &&   + (1-q) \xi^2  | 0 \rangle \langle 0 | \otimes \sum^{d-1}_{j=1}  |j \rangle \langle j | + \bigg( \frac{(1-q)\xi^4 }{q}  + ( \xi^2-\sqrt{q} )^2\bigg) | 0 0 \rangle \langle 0 0 | \bigg]
\end{eqnarray}
\end{widetext}
where, $N_d = [q+ (1-q)\xi^2] (1-\frac{1}{d}) + \frac{ \xi^4}{q d} $ is the normalization factor. Using quantum Fourier transformation \cite{Nielsen2010QuantumCA} for $|k \rangle_{A_a}$ and $|l \rangle_{B_b}$ in Eq.(\ref{ma}) and Eq.(\ref{mb}) for this filtered state, we derive the joint probabilities  as
\begin{eqnarray}
\label{stt}
&&P^F_{Q}(A_a =k , B_b = l)\\ \nonumber && = \frac{1}{d^3 N_d}  \bigg[ \frac{q}{2 \sin^2{[\pi(k-l + \alpha_a +\beta_b)/d]}}   \\ \nonumber &&+ \frac{ \sqrt{q}( \xi^2-  \sqrt{q})}{ \sin{[\pi(k-l + \alpha_a +\beta_b)/d]}}   +S_1 \bigg]
\end{eqnarray}
where, $S_1 =  (1-q)(d-1) \xi^2 +  \frac{(1-q)\xi^4 }{q}  + ( \xi^2-\sqrt{q} )^2$. The detailed derivations of Eq. (\ref{dfr}) and Eq. (\ref{stt}) are placed in Appendix A. Since $S_1$ is independent of $k$, all terms in the joint probabilities follow the symmetric relation given by
\begin{eqnarray}
\label{sym}
P^F_{Q}(A_a =k , B_b = l) = P^F_{Q}(A_a =k+c , B_b = l+c)
\end{eqnarray}
where $c$ is an integer. Using the symmetry in Eq. (\ref{sym}), the optimal quantum value of CGLMP functional is derived as 
\begin{eqnarray}
\label{cgql}
I^F_{d,Q} &=&  4 d \sum^{\left\lfloor \frac{d}{2}\right\rfloor -1}_{k = 0} \bigg( 1 - \frac{2 k}{d-1}     \bigg) (p^F_k - p^F_{-(k+1)} )
\end{eqnarray}

\begin{figure}[htp]
  \includegraphics[width=8cm]{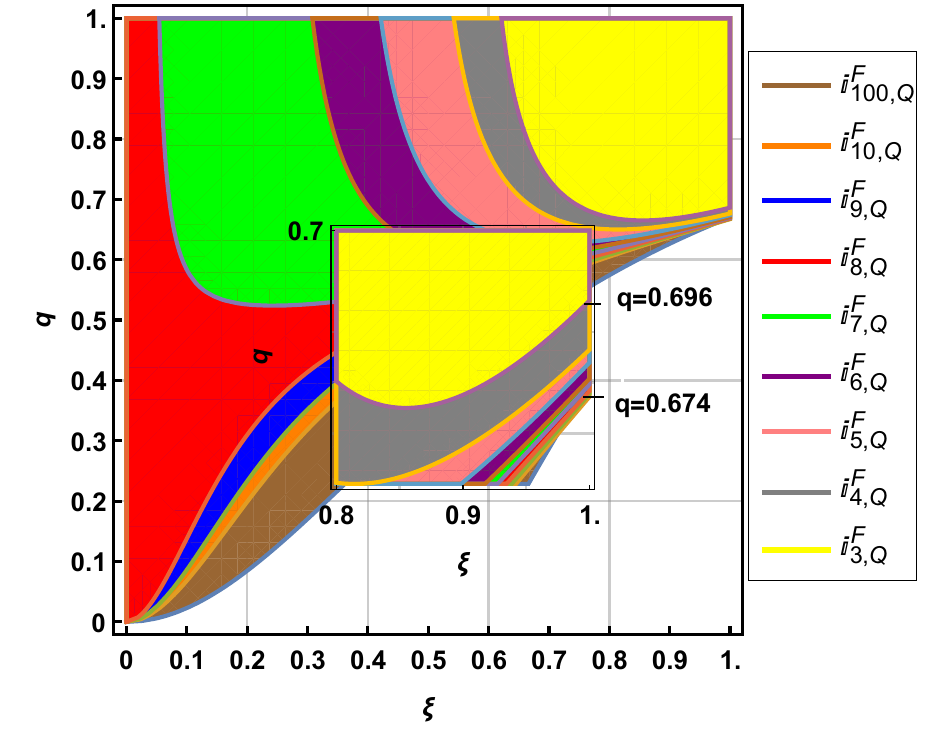}
\caption{Shaded region formed by $I^F_{d,Q} > 2$ shows the violation of CGLMP inequality for $d=3,4,5,6,7,8,9,10$ and $d=100$ after local filtering operation.}
    \label{fig:BLP}
\end{figure}

\begin{figure}[htp]
  \includegraphics[width=8cm]{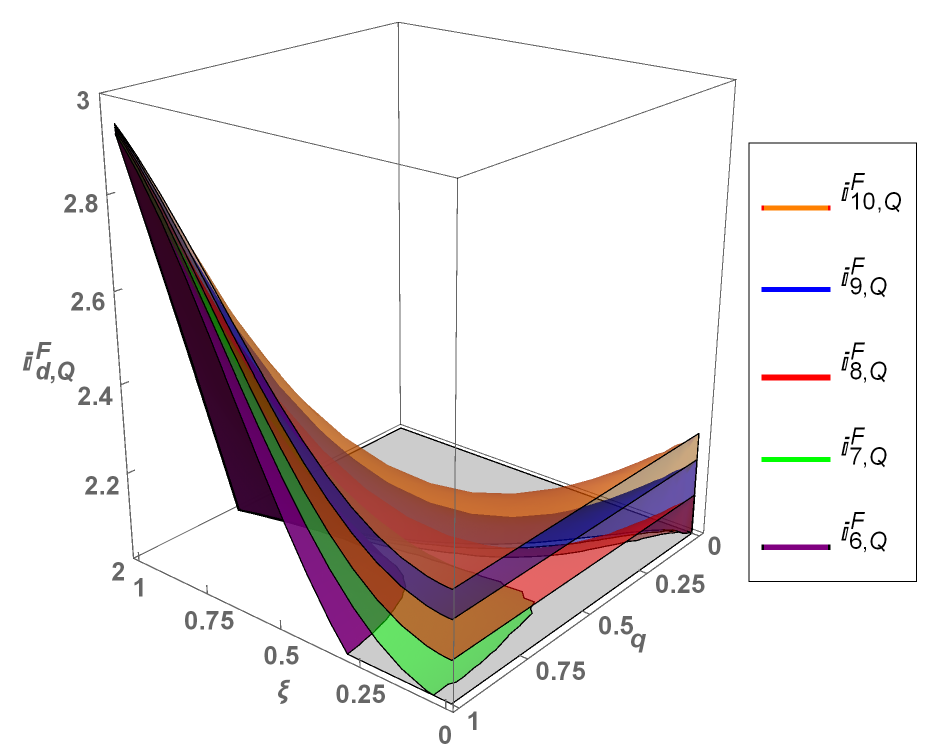}
\caption{Maximum quantum violation of CGLMP inequality for $d=6,7,8,9$ and $d=10$ are plotted with respect to $\xi$ and $q$.}
    \label{fig:BLP1}
\end{figure}
where,
\begin{eqnarray}
\label{jpf}
\nonumber
p^F_c = P^F_{Q}(A_1 = B_1+ c ) &=& \frac{1}{d^3 N_d}  \bigg[ \frac{q}{2 \sin^2{[\pi(c +1/4)/d]}}   \\ &+& \frac{ \sqrt{q}( \xi^2-  \sqrt{q})}{ \sin{[\pi(c + 1/4)/d]}}   +S_1 \Bigg]
\end{eqnarray}

\begin{table}[]
\begin{center}
\begin{tabular}
 {|p{0.5cm}|p{2.0cm}|p{2.0cm}|p{1.9cm}|p{1.0cm}|}  
%\hline
%\multicolumn{4}{|c|}{Optimal quantum value of CGLMP inequality} \\
\hline
\centering $d$ & \centering Nonlocal region before filtering &\centering Nonlocal region after filtering & \centering Region of hidden nonlocality & \hspace{0.35 cm} $\xi$    \\
\hline
\centering $3$ & \centering $0.696 < q \leq 1$ & \centering $0.664 < q \leq 1$ &  $0.664-0.696$ & \hspace{0.25 cm}  $ 0.85 $  \\
\hline
\centering $4$ & \centering $0.690 < q \leq 1$ & \centering $0.648 < q \leq 1$ &  $0.648-0.690$& \hspace{0.25 cm}  $ 0.81$  \\
\hline
\centering $5$ & \centering $0.687 < q \leq 1$ & \centering $0.627 < q \leq 1$ &  $0.627-0.687$ & \hspace{0.25 cm}  $ 0.71 $ \\
\hline
\centering $6$ & \centering $0.684 < q \leq 1$ & \centering $0.610 < q \leq 1$ &  $0.610-0.684$ & \hspace{0.25 cm}  $0.60 $  \\
\hline
\centering $7$ & \centering $0.683 < q \leq 1$ & \centering $0.524 < q \leq 1$ &  $0.524-0.683$& \hspace{0.25 cm}  $0.25 $  \\
\hline
\centering $8$ & \centering $0.682 < q \leq 1$ & \centering $0 < q \leq 1$ &    \hspace{0.25 cm} $0-0.682$ & \hspace{0.25 cm}  $ \rightarrow 0$ \\
\hline
\centering $9$ & \centering $0.681 < q \leq 1$ & \centering $0 < q \leq 1$ &   \hspace{0.25 cm}  $0-0.681$ & \hspace{0.25 cm}  $ \rightarrow 0$ \\
\hline
\centering $10$ & \centering $0.680 < q \leq 1$ & \centering $0 < q \leq 1$ &   \hspace{0.25 cm}  $0-0.680$ & \hspace{0.25 cm}  $ \rightarrow 0$ \\
\hline
\centering $100$ & \centering $0.674 < q \leq 1$ & \centering $0 < q \leq 1$ &   \hspace{0.25 cm}  $0-0.674$ & \hspace{0.25 cm}  $ \rightarrow 0$ \\

\hline
\end{tabular}
\end{center}
\caption{Range of mixing parameter $q$ violating CGLMP inequality before and after local filtering operation when pure state $|\psi_d\rangle$ in Eq. (\ref{pd}) is a maximally entangled state.
   }
 \label{table:max}
\end{table}
 Note that for $\xi =\delta =1$, the local filters in Eq.(\ref{df}) becomes $\mathbb{I}_{d}$, and $I^F_{d,Q}$ reduce to $I'_{d,Q}$ given in Eq.(\ref{cgll}). 
 
It is known from Eq. (\ref{cgll1}) that in obtaining the nonlocality for the state $\rho_{d}$ in Eq. (\ref{pd}), the range of mixing parameters slowly increases with the increment of dimension $d$. For example, for $d=3$ the range $0.696 < q \leq 1$ becomes $0.674 < q \leq 1$ for $d=100$. We demonstrate that by the action of local filtering operations the range of $q$ gradually increases with the increment of $d$ and the decrement of $\xi$. Interestingly, the nonlocality can be revealed for any nonzero value of mixing parameter $q$ for $d\geq 8$ when $\xi\rightarrow 0$. The ranges of mixing parameter $q$ revealing hidden nonlocality for the dimensions for $d=3,4,5,6,7,8,9,10$ and $d=100$ are given in Table {\ref{table:max}}.  

In Fig. 1, we exhibit the nonlocal region i.e., the region satisfies $I^F_{d, Q} > 2$ (CGLMP inequality violation) for various dimensions against the mixing parameter $q$ and the parameter $\xi$ involved in local filter operators. We note that the violation of CGLMP inequality is not obtained for all values of $\xi$. In Fig.2, we show that the quantum violation of CGLMP inequality after local filtering operation for $d=6,7,8,9$ and $d=10$ with respect to $\xi$ and $q$. It shows that local filtering operation does not increase the maximum quantum value of CGLMP inequality but only increases the range of mixing parameter revealing the quantum violation. The range of mixing parameter increases for lower value of quantum violation of CGLMP inequality.

 \subsection{ When $|\psi_d\rangle$  is maximally violating state}

Note that CGLMP inequality is maximally violated by a non-maximally entangled state (known as maximally CGLMP-violating state) if $d > 2$. We consider the same local filtering operations as the previous case. We discuss the range of mixing parameter of $\rho_d$ in Eq.(\ref{pd}) for $d=3,4$ and $d=5$ both before and after local filtering operation.

 The CGLMP inequality in Eq.(\ref{cgl}) for $d=3$, reduces to
\begin{eqnarray}
\label{cg3}
(I_{3})_{L} &=& P(A_1 = B_1) + P(B_1 = A_2+1) +P(A_2 = B_2) \\ \nonumber &+&  P(B_2 = A_1)
-\big(P(A_1 = B_1-1) + P(B_1 = A_2) \\ \nonumber &+& P(A_2 = B_2-1)+P(B_2 = A_1-1)\big) \leq 2
\end{eqnarray}
If the mixed entangled state shared between Alice and Bob is of the form
\begin{eqnarray}
\label{3mv}
\rho_{3} = q |\psi_{3} \rangle \langle \psi_{3} | + (1-q) |0 \rangle \langle 0 | \otimes \frac{\mathbb{I}_3}{3}
\end{eqnarray}
where,
\begin{eqnarray}
\label{p}
|\psi_{3} \rangle =   \gamma_1 |00 \rangle + \gamma_2 |11 \rangle +  \gamma_3 |22 \rangle
\end{eqnarray}
 is maximally violating state for suitable value of $\gamma_1, \gamma_2 $ and $\gamma_3=\sqrt{1-(\gamma_{1}^{2}+\gamma_{2}^{2})} $. For this shared state and operators  $A_a$ ($a = 1, 2$) and $B_b$ ($b = 1, 2$) with eigenvectors given in Eq.(\ref{ma}) and Eq.(\ref{mb}), the maximum quantum value of CGLMP inequality is given by
\begin{eqnarray}
 I_{3,Q} =2.915 q
\end{eqnarray}
obtained at  $\gamma_1 =0.6169, \gamma_2 =0.4888$ and $\gamma_3 =0.6169$ \cite{fonseca18,roy20}. The violation of  the inequality in Eq. (\ref{cg3}) is obtained within the range $0.686 < q \leq 1$. 

To reveal hidden nonlocality outside $0.686 < q \leq 1$, we apply  local filtering operations defined by
\begin{eqnarray}
F_A = \xi | 0 \rangle \langle 0 | +  | 1 \rangle \langle 1 | +  | 2 \rangle \langle 2 | \\ \nonumber
F_B = \delta | 0 \rangle \langle 0 | +  | 1 \rangle \langle 1 | +  | 2 \rangle \langle 2 | \nonumber
\end{eqnarray}
on the respective local part of the shared state where $\delta = \frac{\xi}{\sqrt{q}}$. The quantum value of $(I_{3})_L $ for the filtered state 
\begin{eqnarray}
\rho^F_{3} = \frac{(F_A  \otimes F_B)\rho_{3} (F_A  \otimes F_B)^{\dagger}}{Tr[(F_A  \otimes F_B)\rho_{3} (F_A  \otimes F_B)^{\dagger}]}
\end{eqnarray}
and measuring operators  $A_a$ ($a = 1, 2$) and $B_b$ ($b = 1, 2$) with eigenvectors given in Eq.(\ref{ma}) and Eq.(\ref{mb}) is derived as
\begin{eqnarray}
 (I^F_{3})_{Q} =\frac{2.218 \xi ^2 \sqrt{q}+0.696 q}{\frac{\xi ^4 (0.047 q+0.333)}{q}-0.666 \xi ^2 (q-1.)+0.619 q}
\end{eqnarray}

\begin{table}[]
\begin{center}
\begin{tabular}
 {|p{0.5cm}|p{2.0cm}|p{2.0cm}|p{1.9cm}|p{1.0cm}|}  
%\hline
%\multicolumn{4}{|c|}{Optimal quantum value of CGLMP inequality} \\
\hline
\centering $d$ & \centering  Non-local region before filtering & \centering Non-local region after filtering & \centering Region of hidden non-locality & \hspace{0.35 cm} $\xi$    \\
\hline
\centering $3$ & \centering $0.686 < q \leq 1$ & \centering $0.625 < q \leq 1$ &  $0.625-0.686$ & \hspace{0.25 cm}  $ 0.73 $  \\
\hline
\centering $4$ & \centering $0.672 < q \leq 1$ & \centering $0.585 < q \leq 1$ &  $0.585-0.672$& \hspace{0.25 cm}  $ 0.64$  \\
\hline
\centering $5$ & \centering $0.663 < q \leq 1$ & \centering $0.539 < q \leq 1$ &  $0.539-0.663$ & \hspace{0.25 cm}  $ 0.54 $ \\
\hline
\end{tabular}
\end{center}
\caption{Range of mixing parameter $q$ violating CGLMP inequality before and after filtering operation when pure state $|\psi_d\rangle$ in Eq. (\ref{pd}) is a maximally CGLMP violating state.}
 \label{table:nonmax}
\end{table}
We obtain the quantum violation of CGLMP ($(I^F_{3})_Q>2$) for the range of $0.625 < q \leq 1$ at $\xi = 0.73$. This range is wider than the range $0.664 < q \leq 1$ obtained for $\rho_d$ with $|\psi_d \rangle$ as a maximally entangled state. Hence, the range of mixing parameter showing hidden nonlocality using a maximally violating state is $0.625 < q < 0.686 $. 

Further, we derive the range of the mixing parameter for $d=4$ and $5$ as given in Table {\ref{table:nonmax}}. It is seen that the hidden nonlocality can be demonstrated for the lower value of the mixing parameter if the dimension of the system is increased. The detailed derivation of the range of $q$ revealing hidden nonlocality for $d=4$ and $5$ is placed in Appendix B. 

\section{Discussion} 
In summary, we have demonstrated the hidden nonlocality for a class of local entangled states in an arbitrary $d$-dimensional system by using local filtering operations. Popescu's original proposal \cite{Popescu1995} of revealing hidden nonlocality based on the quantum violation of CHSH inequality was demonstrated for Werner state in $d \geq 5$.  We have demonstrated the activation of nonlocality for a class of local entangled state in arbitrary $d$ dimensions using local filtering operation through the quantum violation of CGLMP inequality which has hitherto not been studied.  For this, we have considered a class of local mixed entangled states which is a convex mixture of a pure entangled state and the color noise.

We considered two different cases, viz., when the pure state $|\psi_{d}\rangle$ in Eq. (\ref{pd}) is a maximally entangled state and the maximally CGLMP-violating state. In the former case, we demonstrated that by suitable local filtering operations, the range of mixing parameter $q$ revealing the nonlocality increases with the increment of the dimension of the system. Importantly, for $d \geq 8$, the quantum violation of CGLMP inequality is obtained for any nonzero value of the mixing parameter $q$. In the latter case when the pure state $|\psi_{d}\rangle$ in Eq. (\ref{pd})  is the maximally CGLMP-violating state, we showed that the hidden nonlocality can be demonstrated for a comparatively wider range of $q$ compared to the maximally entangled state for $d=3,4,5$. However, we have shown this case up to $d=5$ and we conjecture that for the dimension $d\geq 6,7$ the range will be wider compared to the former case which can be further studied. 

Finally, we propose a few problems as a follow-up of our study. Note that, our work is restricted to projective measurement and hence generalizing it for POVMs could be an interesting line of future research. We have considered a specific class of mixed entangled state admitting local models. It will then be interesting to study the activation of nonlocality for other class of mixed entangled states through the quantum violation of CGLMP inequality. This could also be an interesting avenue for future research which calls for further study.

\section*{Acknowledgments}
 Asmita Kumari acknowledges the Research Associateship from S. N. Bose National Centre for Basic Sciences, Kolkata, India.

\begin{widetext}
\appendix
\section{Hidden nonlocality in $d$-dimensional mixed entangled state.}
To investigate hidden nonlocality for the state  $\rho_{d}$ in Eq.(\ref{pd}) with $|\psi_d \rangle$ as maximally entangled state, let us assume that Alice and Bob perform local filter operations defined in Eq.(\ref{df})
on their respective local part of the shared state. The operation of local filters transform the shared state $\rho_{d}$ to un-normalized state given by
\begin{eqnarray}
\label{a}
\Tilde{\rho}^F_{d} &=&  (F_A \otimes F_B) \rho_{d}  (F_A \otimes F_B)^{\dagger} \\ \nonumber &=& q (F_A \otimes F_B) |\psi_d \rangle \langle \psi_d |  (F_A \otimes F_B)^{\dagger} + \frac{1-q}{d} (F_A \otimes F_B) | 0 \rangle \langle 0 | \otimes \mathbb{I}_d  (F_A \otimes F_B)^{\dagger} \\ \nonumber &=&   \frac{q}{d}\bigg[ \delta \xi | 0 0 \rangle +  \sum^{d-1}_{j=1}  |j \rangle_A   |j \rangle_B \bigg]\bigg[ \delta \xi \langle 0 0 | +  \sum^{d-1}_{j=1}   \langle j |  \langle  j | \bigg] +  \bigg( \frac{1-q}{d} \bigg)  \xi^2  | 0 \rangle \langle 0 | \otimes \bigg( \delta^2 | 0 \rangle \langle 0 | +  \sum^{d-1}_{j=1}  |j \rangle \langle  j|  \bigg)  \\ \nonumber  &=&   \frac{q}{d}\bigg[ (\delta \xi -1 )| 0 0 \rangle +  \sum^{d-1}_{j=0}  |j \rangle   |j \rangle \bigg]\bigg[ (\delta \xi-1 ) \langle 0 0 | +  \sum^{d-1}_{j=0}   \langle j |  \langle  j | \bigg] + \bigg( \frac{1-q}{d} \bigg)  \xi^2  | 0 \rangle \langle 0 | \otimes \bigg( \delta^2 | 0 \rangle \langle 0 | +  \sum^{d-1}_{j=1}  |j \rangle \langle  j |  \bigg) \\ \nonumber  &=& \frac{1}{d} \bigg[ q(\delta \xi-1 )^2 | 0 0 \rangle \langle 0 0 |+ q(\delta \xi-1 ) \bigg( \sum^{d-1}_{j=0}  |j \rangle   |j \rangle \bigg) \langle 0 0 | + q(\delta \xi-1 ) | 0 0 \rangle  \bigg( \sum^{d-1}_{j=0}   \langle j |  \langle  j | \bigg)   \\ \nonumber &&  + q \bigg( \sum^{d-1}_{j=0}  |j \rangle   |j \rangle \bigg) \bigg( \sum^{d-1}_{j=0}   \langle j |  \langle  j | \bigg) +  (1-q) \delta^2 \xi^2 | 0 \rangle \langle 0 | \otimes  | 0 \rangle \langle 0 | + (1-q) \xi^2  | 0 \rangle \langle 0 | \otimes \sum^{d-1}_{j=1}  |j \rangle \langle j |  \bigg]
\end{eqnarray}
The normalized filtered state with $\delta =\frac{\xi }{\sqrt{q}}$ can be re-written as
\begin{eqnarray}
\nonumber
\rho^F_{d}   &=&  \frac{1}{d  N_d} \bigg[q \bigg( \sum^{d-1}_{j=0}  |j \rangle   |j \rangle \bigg) \bigg( \sum^{d-1}_{j=0}   \langle j |  \langle  j | \bigg)+ \sqrt{q}( \xi^2-  \sqrt{q}) \bigg[ \bigg( \sum^{d-1}_{j=0}  |j \rangle   |j \rangle \bigg) \langle 0 0 | +| 0 0 \rangle \bigg( \sum^{d-1}_{j=0}   \langle j |  \langle  j |  \bigg) \bigg]  \\ \nonumber &&   + (1-q) \xi^2  | 0 \rangle \langle 0 | \otimes \sum^{d-1}_{j=1}  |j \rangle \langle j | + \bigg( \frac{(1-q)\xi^4 }{q}  + ( \xi^2-\sqrt{q} )^2\bigg) | 0 0 \rangle \langle 0 0 | \bigg]
\end{eqnarray}

Where, $N_d = [q+ (1-q)\xi^2] (1-\frac{1}{d}) + \frac{ \xi^4}{q d} $ is the normalization constant. The next task of Alice and Bob is to perform local measurements using operators  $A_a$ ($a = 1, 2$) and $B_b$ ($b = 1, 2$) with eigenvectors $|k \rangle_{A_a}$ (Eq.(\ref{ma})) and $|l \rangle_{B_b}$ (Eq.(\ref{mb})) on their respective part of the shared filtered state $\rho^F_{d} $. The joint correlation function for this filtered state can be calculated as given below.
\begin{eqnarray}
P^F_{QM}(A_a =k , B_b = l) &=& Tr\bigg[ ( |k \rangle_{A_a} \langle k| \otimes  |l \rangle_{B_b} \langle l|)   \rho^F_{d} \bigg]  \\ \nonumber  &=& \frac{1}{d N_d} \bigg[q Tr\bigg[ ( |k \rangle_{A_a} \langle k| \otimes  |l \rangle_{B_b} \langle l|)  \bigg( \sum^{d-1}_{j=0}  |j \rangle   |j \rangle \bigg) \bigg( \sum^{d-1}_{j=0}   \langle j |  \langle  j | \bigg) \bigg] \\ \nonumber && +\sqrt{q}( \xi^2-  \sqrt{q}) Tr\bigg[ ( |k \rangle_{A_a} \langle k| \otimes  |l \rangle_{B_b} \langle l|)  \bigg( \bigg( \sum^{d-1}_{j=0}  |j \rangle   |j \rangle \bigg)\langle 0 0 | +| 0 0 \rangle \bigg( \sum^{d-1}_{j=0}   \langle j |  \langle  j |  \bigg) \bigg) \bigg] \\ \nonumber && +  (1-q) \xi^2 Tr\bigg[  ( |k \rangle_{A_a} \langle k| \otimes  |l \rangle_{B_b} \langle l|) | 0 \rangle \langle 0 | \otimes \sum^{d-1}_{j=1}  |j \rangle \langle j |\bigg] \\ \nonumber && + \bigg( \frac{(1-q)\xi^4 }{q}  + ( \xi^2-\sqrt{q} )^2\bigg)Tr\bigg[ ( |k \rangle_{A_a} \langle k| \otimes  |l \rangle_{B_b} \langle l|) | 0 0 \rangle \langle 0 0 | \bigg]\bigg] 
\end{eqnarray}
Except $q$ factor, first term of $P^F_{QM}(A_a =k , B_b = l)$ is same as the joint probability $P_{QM}(A_a =k , B_b = l)$ of unfiltered state. Using the quantum Fourier transformation \cite{Nielsen2010QuantumCA} for  $|k \rangle_{A_a}$ and $ |l \rangle_{B_b} $ for the second term, the joint probability $P^F_{QM}(A_a =k , B_b = l)$ can be re-written as
\begin{eqnarray}
P^F_{QM}(A_a =k , B_b = l) &=&  \frac{1}{d^3 N_d}  \bigg[ q  \left| \sum^{d-1}_{j=0} \exp{\bigg[ i \frac{2 \pi j}{d}(k-l + \alpha_a +\beta_b) \bigg]} \right|^2 \\ \nonumber && + \sqrt{q}( \xi^2-  \sqrt{q}) \bigg(  \sum^{d-1}_{j=0}\exp{\bigg[ - i \frac{2 \pi j}{d}(k-l + \alpha_a +\beta_b) \bigg]} + \sum^{d-1}_{j=0} \exp{\bigg[ i \frac{2 \pi j}{d}(k-l + \alpha_a +\beta_b) \bigg]}   \bigg) \\ \nonumber &&  + (1-q)(d-1) \xi^2 + \bigg( \frac{(1-q)\xi^4 }{q}  + ( \xi^2-\sqrt{q} )^2\bigg) \bigg] \\ \nonumber & = & \frac{1}{d^3 N_d}  \bigg[ \frac{q\sin^2{[\pi(k-l + \alpha_a +\beta_b)]}}{ \sin^2{[\pi(k-l + \alpha_a +\beta_b)/d]}} + \frac{ \sqrt{q}( \xi^2-  \sqrt{q})\sin{[2\pi(k-l + \alpha_a +\beta_b)]}}{ \sin{[\pi(k-l + \alpha_a +\beta_b)/d]}}  \\ \nonumber &&  + (1-q)(d-1) \xi^2 +  \frac{(1-q)\xi^4 }{q}  + ( \xi^2-\sqrt{q} )^2 \bigg]
\end{eqnarray}

Last line is obtained using $\sum^{M-1}_{m=0} \exp{(imx)} = \frac{\sin{(Mx/2)}}{\sin{(x/2)}}\exp{(ix(M-1)/2)}$. Substituting the values of $\alpha_a$ and $\beta_b$ in Eq(\ref{v1}), we have
\begin{eqnarray}
P^F_{QM}(A_a =k , B_b = l) &=& \frac{1}{d^3 N_d}  \bigg[ \frac{q}{2 \sin^2{[\pi(k-l + \alpha_a +\beta_b)/d]}} + \frac{ \sqrt{q}( \xi^2-  \sqrt{q})}{ \sin{[\pi(k-l + \alpha_a +\beta_b)/d]}}  \\ \nonumber &&  + (1-q)(d-1) \xi^2 +  \frac{(1-q)\xi^4 }{q}  + ( \xi^2-\sqrt{q} )^2 \bigg]
\end{eqnarray}
 Using this joint probability the optimal quantum value of CGLMP functional after local filtering operation can be obtained.\\

\section{When $|\psi_d\rangle$  is maximally violating state}

\subsection{{Detailed calculation for $d=4$}}

The CGLMP inequality for $d=4$ is given by
\begin{eqnarray}
\label{cg4}
(I_{4})_L &=& P(A_1 = B_1) + P(B_1 = A_2+1) +P(A_2 = B_2) + P(B_2 = A_1) \\ \nonumber & -& (P(A_1 = B_1-1) + P(B_1 = A_2)+P(A_2 = B_2-1)+P(B_2 = A_1-1)) \\ \nonumber &+&\frac{1}{3}\bigg(P(A_1 = B_1 + 1) +P(B_1 = A_2+2)+ P(A_2 = B_2 + 1) +P(B_2 = A_1 + 1 ) \\ \nonumber & -&(P(A_1 = B_1-2) P(B_1 = A_2 - 1)+  P(A_2 = B_2-2) +P(B_2 = A_1-2))\bigg)\leq 2.
\end{eqnarray}
We take shared entangled states of the form
\begin{eqnarray}
\label{p}
\rho_{4} = q |\psi_{4} \rangle \langle \psi_{4} | + (1-q) |0 \rangle \langle 0 | \otimes \frac{\mathbb{I}_4}{4}
\end{eqnarray}
where
\begin{eqnarray}
\label{p}
|\psi_{4} \rangle =  \gamma_1 |00 \rangle + \gamma_2 |11 \rangle +  \gamma_3 |22 \rangle +  \gamma_4 |33 \rangle
\end{eqnarray}
is the maximally violating state for suitable value of $\gamma_1, \gamma_2, \gamma_3 $ and $\gamma_4=\sqrt{1-(\gamma_{1}^{2}+\gamma_{2}^{2}+\gamma_{3}^{2})} $.
For this shared state and operators  $A_a$ ($a = 1, 2$) and $B_b$ ($b = 1, 2$) with eigenvectors given in Eq.(\ref{ma}) and Eq.(\ref{mb}), we derive the maximum quantum value  of CGLMP inequality $(I_{4})_L$ as
\begin{eqnarray}
 I_{4,Q} = 2.972 q
\end{eqnarray}
at $\gamma_1 =0.5686, \gamma_2 =0.4204, \gamma_3 =0.4204$, and $\gamma_4 =0.5686$ \cite{fonseca18,roy20}. The state $|\psi_{4} \rangle$ is a maximally violating state with these values of coefficient.
 We get  $(I_{4})_L >2$ when $0.672 < q \leq 1$ and hence the state $\rho_4$ is local in the range of $0 < q \leq 0.672$.

In order to reveal hidden nonlocality in $0 < q \leq 0.672$, let us apply a local filtering operation defined by
\begin{eqnarray}
\label{f4}
F_A = \xi | 0 \rangle \langle 0 | +  | 1 \rangle \langle 1 | +  | 2 \rangle \langle 2 | +  | 3 \rangle \langle 3 | \\ \nonumber
F_B = \delta | 0 \rangle \langle 0 | +  | 1 \rangle \langle 1 | +  | 2 \rangle \langle 2 | +  | 3 \rangle \langle 3 |\\ \nonumber
\end{eqnarray}
on their respective part of the shared state. The quantum value of $(I_{4})_L $ for the filtered state 
\begin{eqnarray}
\rho^F_{4} = \frac{(F_A  \otimes F_B)\rho_{4} (F_A  \otimes F_B)^{\dagger}}{Tr[(F_A  \otimes F_B)\rho_{4} (F_A  \otimes F_B)^{\dagger}]}
\end{eqnarray}
and the measurement settings in Eq.(\ref{ma}) and Eq.(\ref{mb}) is derived as 
\begin{eqnarray}
 (I^F_{4})_{Q} =\frac{-2.562 \xi ^2 q^{3/2}-1.401 q^2}{-0.902 q^2+\xi ^4 (-0.097 q-0.333)-\xi ^2 q (1-q)}
\end{eqnarray}
In this case we obtain quantum violation of CGLMP inequality for the range of $0.585 < q \leq 1$ at $\xi = 0.64$,  which is wider in comparison to $0.648 < q \leq 1$ obtained for $\rho_4$ with $|\psi_4 \rangle$ as a maximally entangled state.

\subsection{{Detailed calculation for $d=5$}}

Substituting $d = 5$ in Eq.(\ref{cgl}), the CGLMP inequality reduces to
\begin{eqnarray}
\label{cg5}
(I_{5})_L &=& P(A_1 = B_1) + P(B_1 = A_2+1) +P(A_2 = B_2)+ P(B_2 = A_1)  \\ \nonumber &-&  [P(A_1 = B_1-1) + P(B_1 = A_2) +P(A_2 = B_2-1)+P(B_2 = A_1-1)] \\ \nonumber &+&  \frac{1}{2}\bigg(P(A_1 = B_1 + 1) +P(B_1 = A_2+2)+ P(A_2 = B_2 + 1) +P(B_2 = A_1 + 1 ) \\ \nonumber &-& [P(A_1 = B_1-2) P(B_1 = A_2 - 1) + P(A_2 = B_2-2) +P(B_2 = A_1-2)]\bigg)\leq 2 
\end{eqnarray}
We take the mixed entangled state 
\begin{eqnarray}
\label{p}
\rho_{5} = q |\psi_{5} \rangle \langle \psi_{5} | + (1-q) |0 \rangle \langle 0 |\frac{\mathbb{I}_5}{5}
\end{eqnarray}
where,
\begin{eqnarray}
\label{p}
|\psi_{5} \rangle =  \gamma_1 |00 \rangle + \gamma_2 |11 \rangle +  \gamma_3 |22 \rangle +  \gamma_4 |33 \rangle+  \gamma_5 |44 \rangle
\end{eqnarray}
is maximally violating state for suitable value of $\gamma_1, \gamma_2, \gamma_3, \gamma_4$ and $\gamma_5=\sqrt{1-(\gamma_{1}^{2}+\gamma_{2}^{2}+\gamma_{3}^{2}+\gamma_{4}^{2})} $.
For the shared state $|\psi_{5} \rangle$ and operators  $A_a$ ($a = 1, 2$) and $B_b$ ($b = 1, 2$) with eigenvectors given in Eq.(\ref{ma}) and Eq.(\ref{mb}), the maximum quantum value  of CGLMP inequality  $(I_{5})_L$ is 
\begin{eqnarray}
 I_{5,Q} =3.0158 q
\end{eqnarray}
obtained at $\gamma_1 =0.5368, \gamma_2 =0.3859, \gamma_3 =0.3548$, and $\gamma_4 =0.3859$ and $\gamma_5 =0.5368$  \cite{fonseca18,roy20}.
In this case, nonlocality is observed in the range of $0.663 < q \leq 1$, and hence $\rho_5$ is local in the range of $0 < q \leq 0.663$.

In order to reveal hidden nonlocality in $0 < q \leq 0.663$, we consider the local filtering operations defined by
\begin{eqnarray}
\label{f5}
F_A = \xi | 0 \rangle \langle 0 | +  | 1 \rangle \langle 1 | +  | 2 \rangle \langle 2 | +  | 3 \rangle \langle 3 |+  | 4 \rangle \langle 4 | \\ \nonumber
F_B = \delta | 0 \rangle \langle 0 | +  | 1 \rangle \langle 1 | +  | 2 \rangle \langle 2 | +  | 3 \rangle \langle 3 |+  | 4 \rangle \langle 4 |\\ \nonumber
\end{eqnarray}
on their respective part of the shared state. The quantum value of $(I_{5})_L $ for the filtered state
\begin{eqnarray}
\rho^F_{5} = \frac{(F_A  \otimes F_B)\rho_{5} (F_A  \otimes F_B)^{\dagger}}{Tr[(F_A  \otimes F_B)\rho_{5} (F_A  \otimes F_B)^{\dagger}]}
\end{eqnarray}
and the measuring operators  $A_a$ ($a = 1, 2$) and $B_b$ ($b = 1, 2$) with eigenvectors given in Eq.(\ref{ma}) and Eq.(\ref{mb}) is given by
\begin{eqnarray}
 (I^F_{5})_{Q} = \frac{-2.172 \xi ^2 q^{3/2}-1.597 q^2}{-0.889 q^2+\xi ^4 (-0.110 q-0.25)-\xi ^2 q (1-q)}
\end{eqnarray}
We obtain the quantum violation of CGLMP inequality ($(I^F_{5})_Q>2$) for the range of $0.539 < q \leq 1$ at $\xi = 0.54$. This range is wider than the range $0.627 < q \leq 1$ obtained for $\rho_d$ with $|\psi_d \rangle$ as a maximally entangled state. Hence, the range of the mixing parameter showing hidden nonlocality using a maximally violating state is $0.539 < q < 0.627 $.

\end{widetext}

\end{document}